\documentstyle{article}
\begin{document}
\LARGE
\begin{center}
\bf Entropy of a Black Hole with Distinct Surface Gravities 

\vspace*{0.6in}
\normalsize \large \rm

Zhong Chao Wu

Dept. of Physics

Beijing Normal University

Beijing 100875, China

\vspace*{0.4in}
\large
\bf
Abstract
\end{center}
\vspace*{.1in}
\rm
\normalsize
\vspace*{0.1in}

In gravitational thermodynamics, the entropy of a black hole
with distinct surface gravities can be evaluated in a
microcanonical ensemble. At the $WKB$ level,  the entropy 
becomes the negative of the Euclidean action of the constrained
instanton, which is the seed for the black hole creation in the
no-boundary universe. Using the Gauss-Bonnet theorem, we prove
the quite universal formula in Euclidean quantum gravity 
that the entropy of a nonrotating black hole is one quarter
the sum of the products of the Euler characteristics and the
areas of the horizons. For Lovelock gravity, the entropy and
quantum creation of a black hole are also studied.

\vspace*{0.3in}

PACS number(s): 98.80.Hw, 98.80.Bp, 04.60.Kz, 04.70.Dy


\vspace*{0.3in}

e-mail: wu@axp3g9.icra.it

\pagebreak

In gravitational thermodynamics [1][2], it is well known that, if
a black hole has one horizon or two horizons with equal
surface gravities $\kappa_i$, then one can set the hole to
be in 
equilibrium with the temperature $\kappa_i/2\pi$. The
Bekenstein-Hawking entropy can be calculated. It is determined by
the Euler characteristics and areas of the horizons [3]. It is
widely 
believed, but not proven yet, that the entropy of a generic black
hole,
i.e. a black hole with different surface gravities or
temperatures at the
associated horizons, is one quarter the sum of the products of
the 
Euler characteristics  and the areas of the horizons.
The fact that there does not exist an uniform
temperature for this configuration implies that  the canonical or
grandcanonical ensembles do not apply. Fortunately, one can
circumvent this obstacle using microcanonical ensemble. In
contrast, for microcanonical ensembles, the temperature is not
defined, but all conserved quantities are given. Since the
probability of each state
under the conserved quantity constraints are equal, the entropy
is simply the
logarithm of the number of these states. One can also extend this
argument to such an extreme extent that, for a system without any
conserved quantity fixed and all states are equally probable,
then the entropy becomes the logarithm of the total number of the
states. Under this situation one should obtain the maximum
entropy. In the following, this ensemble is called 
supercanonical ensemble, for convenience.
 
The partition function in gravitational thermodynamics is the
path integral
\begin{equation}
Z = \int d[g_{\mu \nu}]d [\phi]\exp-I
\end{equation}
where the path integral is over all spacetime metrics $g_{\mu
\nu}$ and matter configurations $\phi$ under the conditions
corresponding to
these ensembles, and $I$ is the Euclidean action. The main
contribution to the path integral is
from a stationary action orbit. The partition function is
approximated by the exponential of the negative of the action of
the orbit.
This is called the $WKB$ approximation, which we shall use
in this paper. In the no-boundary universe [4], the stationary
action
orbit is also the seed of the quantum creation of a universe from
nothing. 

If the action of the orbit is stationary with respect to all
variations, then it is called an instanton. It is determined
essentially by the topological properties of the manifold. The
metric is regular without any singularity. For example, the $S^4$
and $S^2 \times S^2$ instantons are the seeds for the de
Sitter and Nariai universes. The Nariai universe is interpreted
as a Schwarzschild-de Sitter black hole with the maximal mass
[5].
The regular instanton is used for the $WKB$ approximation to the
partition function in the supercanonical ensemble.

Recently, it was realized that a regular instanton is the seed of
a universe only for a creation with a stationary probability [6].
In order not to exclude many interesting phenomena, for example,
the quantum creation of a black hole with distinct surface
gravities,
from the study, one has to appeal to the concept of
so-called constrained instanton [7][8]. It is a manifold for
which the action is
stationary with respect only to the variations under some
restrictions, instead to all variations under no restrictions. 

For the quantum creation scenario in the no-boundary universe,
the restrictions are that the 3-metric and matter content on it
of
the created universe are given. These constraints can be
characterized by a few parameters, like mass $m$, charge $Q$ and
angular momentum $J$ for the black hole case. These conditions
are exactly the same for
the microcanonical ensemble in gravitational thermodynamics when
all these parameters are fixed. Therefore, the contribution of
the constrained instanton also dominates the corresponding
partition function. The partition function is the exponential of
the negative of the action of the constrained instanton. 

Since the entropy is the logarithm of the partition function in
both microcanonical and supercanonical ensembles, then at the
$WKB$
level, the entropy is the negative of the action.

The other motivation of this paper is as follows. In a series
of previous papers the problem of quantum creation of Kerr-Newman
black hole families in the de Sitter, anti-de Sitter or Minkowski
space background has been discussed by using constrained
instantons [7][8]. The cases of quantum creations of a
topological black
hole with distinct surface gravities and the $BTZ$ black hole
were also studied [9][10] in the same way. For all cases of
black holes
considered, the
spacetime has a $U(1)$ isometry. The group parameter is
identified as the Killing time coordinate. The constrained
instanton is constructed from a section of the complex version of
the black hole metric by identifying the imaginary time
coordinate $\tau$ with a period $\beta$. The field equations are
obeyed everywhere with the possible exception at  the horizons,
in which a conical singularity may occur. Whether the action of
the pasted manifold is stationary and the manifold is a
constrained instanton only depends on whether the action is
stationary with respect to   $\beta$,
the only degree of freedom left. It has been shown that the
action is indeed independent of $\beta$. However, this was
confirmed case by case in the previous papers. A natural question
arises: is there a deep reason behind this
``coincidence''? The answer is yes.

In this paper we shall use the dimensional continuation of the
Gauss-Bonnet theorem to prove the above longstanding conjecture
on the entropy of a nonrotating black hole with distinct surface
gravities. The byproduct is to show that the origin of the
``coincidence''  in the constrained instanton is due to the
topological implication of the Hilbert-Einstein action, therefore
our argument can be generalized into a wider framework, including
the Lovelock theory of gravitation.

The Euclidean action in the Einstein theory is [1]
\begin{equation}
I = - \frac{1}{16 \pi} \int_M \sqrt{g}(R - 2 \Lambda + L_m) 
d^4x- \frac{1}{8\pi}
\oint_{\partial M }\sqrt{g} Kd^3 x,
\end{equation}
where $R$ is the scalar curvature of the spacetime manifold $M$,
$K$ is the trace of the second form of the boundary $\partial M$,
$g$ is the determinant of the metric for the 4-metric or its
lower dimensional version, $\Lambda$ is the cosmological
constant, and $L_m$ is the Lagrangian of the matter content.

For all nonrotating black hole cases considered, the Euclidean
spacetime metric takes the form
\begin{equation}
ds^2 = \Delta (r) d\tau^2 + \Delta^{-1}(r) dr^2 + r^2
d\Omega^2_2,
\end{equation}
where $\tau = it$, and the 2-metric $d\Omega^2_2$ is a compact
manifold which
does not depend on coordinates $\tau$ and $r$. The zeros of the
rational function $\Delta (r)$ are the horizons. 

One can construct a compact constrained instanton using a sector
between two horizons  denoted by two zeros $r_l, r_k$ in the
identified manifold. The surface gravity $\kappa_i$ of the
horizon $r_i$ is $- d \Delta (r)/2dr| _{r = r_i}$. If the zero is
of multiplicity 1, then one obtains a nonzero $\kappa_i$. On the
two dimensional space $(\tau, r)$ the conical singularity at the
horizon can be regularized by choosing $\beta = 2\pi \kappa_i
^{-1}$.  For two horizons with same nonzero surface gravities,
one can use the same $\beta$ to obtain a compact regular
instanton. If these two surface gravities are different, then the
constrained instanton, which is to be justified later, has at
least one conical singularity at the horizons, since no value of
parameter $\beta$ can regularize the both horizons
simultaneously. The de Sitter model is an exception, since $r$ is
identified with $-r$, only one horizon, i.e. the cosmological
horizon is needed for the construction of the instanton.

If one of the two zeros is of multiplicity larger than 1,
then its surface gravity $\kappa_i = 0$, and the associated
horizon recedes to an internal infinity. Then it is always
possible to regularize the other
horizon by choosing a right value $\beta$ to obtain a regular
instanton. The most familiar case is the extreme
Reissner-Nordstr$\rm\bf\ddot{o}$m black hole in the nonvacuum
model [3].

Now, let us calculate the action of the pasted manifold. We
use $M_l$ to denote the small neighbourhood of horizon $r_l$ with
the boundary of a constant coordinate $r$. The Euler number
$\chi(l)$ for the 2-dimensional $(\tau, r)$ section of
neighbourhood with zero (nonzero) surface gravity is 0 (1). We
use $M^\prime$ to denote $M$ minus $M_l$ and $M_k$. For the form
of action (2), the total action is the sum of those from the
three submanifolds.

First of all, let us consider the vacuum model with 
a cosmological constant. The total action is [11][12]
\begin{equation}
I = I_l + I_k + \int_{M^\prime} (\pi^{ij}\dot{h}_{ij} - NH_0 -N_i
H^i)d^3x d\tau,
\end{equation}
where the actions $I_l$ and $I_k$ are the actions for $M_l$ and
$M_k$. The action of $M^\prime$ has been recast into the
canonical form. $N$ and $N_i$ are
the lapse function and shift vector, $h_{ij}$ and $\pi^{ij}$ are
the 3-metric and the conjugate momenta respectively, $H_0$ and
$H^i$
are the
Einstein and momentum constraints, and the dot denotes the time
derivative. The manifold satisfies the Einstein
equation, and all time derivatives vanish due to the $U(1)$
isometry, therefore the integral over $M^\prime$
is equal zero.

Now the action $I_l$ or $I_k$ can be written
\begin{equation}
I_i =- \frac{1}{16 \pi} \int_{M_i}\sqrt{g} (^4R - 2 \Lambda)d^4x
- \frac{1}{8\pi} \oint_{\partial M_i }\sqrt{g} ^3K d^3x,\;\; (i =
l, k),
\end{equation}
where $^4R$ denotes the 4-dimensional scalar curvature and $^3K$
is the expansion rate of the boundary. If there is a conical
singularity at the horizon, its contribution to the action
can be considered
as the degenerate version of the second term of the action, in
addition to that from the boundary of $M_i$. The conical
singularity 
contribution is termed as a deficit ``angle'' due to its
emergence. If the horizon recedes into an internal infinity, then
this is no longer of concern.

One can apply the Gauss-Bonnet theorem to the 2-dimensional
$(\tau, r)$ section of $M_i$,
\begin{equation}
 \frac{1}{4 \pi} \int_{\hat{M}_i}\sqrt{g}^2Rd^2x +
\frac{1}{2\pi}
\oint_{\partial \hat{M}_i }\sqrt{g} ^1K d^1x +
\frac{\delta_i}{2\pi} = \chi (i),
\end{equation}
where $\hat{M}_i$ is the projection of $M_i$ onto the 
2-dimensional $(\tau, r)$ section, $^2R$ is the scalar curvature
on
it, $^1K$ is the corresponding expansion rate, $\delta_i$ is the
total deficit angle, and
$\chi(i)$ is the Euler characteristic of $\hat{M}_i$. Since the
expansion rate of the
subspace $r^2 d\Omega^2_2$ goes
to zero at the horizon, $^3K$ and $^1K$ are equal. Comparing eqs
(5) and (6), one can see that as the
circumference of the boundary tends to infinitesimal, the action
(5) becomes $-\chi(i) A_i/4$, where $A_i$ is the surface area of
the horizon. It is noted that both the volume integral of (5) and
the first term of the left hand side of (6)  vanish as the
boundary approaches the horizon. This is true even for the
electrovacuum model later. The same result is obtained
regardless of the existence of the conical singularity at the
horizon or not, i.e., it is independent of the value $\beta$.
Here the conical singularity contribution is represented
by $\delta_i/2\pi$.

From (4)-(6), we learn that the action is independent of the
parameter $\beta$.
Therefore, the manifold is qualified as a constrained instanton.
The entropy, or the negative of the total action of the
constrained instanton is
\begin{equation}
S = - I =  \frac{1}{4} ( \chi(l) A_l + \chi(k) A_k).
\end{equation}
This is a quite universal formula.

The Lorentzian universe created from the instanton can be
obtained through a series of analytic continuations from the
equator which halves the instanton. The equator is the joint of
two sections $\tau = const.$ connecting the two horizons. For de
Sitter model, the equator is one section $\tau = const.$ passing
through the cosmological horizon. The
relative creation probability is the exponential of the negative
of the action, i.e., the exponential of the entropy.

The beautiful aspect of the above argument is that we have shown
that the action is totally determined by the topologies and the
areas of the horizons. The origin of this phenomenon is
that the Hilbert-Einstein action is the dimensional continuation
of the Gauss-Bonnet theorem. 

If one use $r_0$ to denote the maximum zero of $\Delta(r)$ and
the metric of the sector $r > r_0$ is Euclidean, then this sector
can also be used for constructing an open instanton. Since the
infinity side is open, one only needs to regularize the conical
singularity for horizon $r_0$. Strictly speaking, the instanton
obtained is constrained, since the action is stationary under
some conditions imposed usually at the infinity, say the
temperature there.

The action of the open instanton is divergent. There are two ways
to regularize this. The first method is called the background
subtraction method [1]. The second method is to use action (4),
where $I_l = I_0$, $M_k$ is dropped and $M^\prime$ is the
sector $M$
minus $M_0$. The action is
\begin{equation}
I = I_0 + \int_{M^\prime} (\pi^{ij}\dot{h}_{ij} - NH_0 -N_i
H^i)d^3x d\tau .
\end{equation}

The form of this action is derived from the
requirement that, as the mass $m$ (and the angular momentum $J$
for a
rotating model not expressed by (3)) is
held fixed at infinity with an appropriate asymptotic falloff
for the field, the Einstein and field equations must be derived
from the action [13]. These boundary conditions correspond to the
microcanonical ensemble. The divergence problem is automatically
cured by this prescription.   Follow the same argument, one can
derive that the entropy is the negative of the action of the
constrained instanton in the microcanonical ensemble, which is
$\chi(0) A_0/4$. The instanton is
constructed by the
pasting method with an arbitrary $\beta$. To avoid the conical
singularity one can choose $\beta = 2\pi \kappa_0^{-1}$, but for
the microcanonical ensemble, one does not have to do so.

This very well explains why the action does depend on $\beta$
when we fix the conjugate of $m$ 
instead of $m$  itself. Then one needs an extra boundary term
$\beta m$ at infinity, and the boundary condition does not
correspond to the microcanonical ensemble. The 
action no longer takes advantage of the dimensional continuation
of the Gauss-Bonnet theorem. The same argument applies to the
angular momentum case.

The method of the dimensional continuation of the Gauss-Bonnet
theorem has been used to study the  entropy of a black hole with
a regular instanton [14][15].

Now we include the Maxwell or Gauge field into the model. One
consider the spatial components of the gauge potential $A_i$ and
the electric gauge field $E^i$ as the canonical coordinates and
conjugate momenta, respectively. The time component $A_\tau$ is
the Lagrange multiplier of the Gauss constraint $\mbox{div} E =
0$. For this case, the action (4) should be revised into the form
[16]
\begin{equation}
I = I_l + I_k + \int_{M^\prime} (\pi^{ij}\dot{h}_{ij} +
E^i\dot{A}_i - NH_0 -N_i
H^i - NA_\tau \mbox{div}E)d^3x d\tau -
\frac{1}{4\pi}\int_{\partial M^\prime}\sqrt{g}A_\tau Ed^3x ,
\end{equation}
where the boundary term at $\partial M^\prime$ is obtained
through integrating by parts the term involving the spatial
derivative of $A_\tau$.

We shall use $Q$ to denote the magnetic or electric charge of the
hole. For the magnetic case, the gauge field is
\begin{equation}
F = Q \cdot \rm (area \;\; element),
\end{equation}
where the area element is that of the metric $d\Omega^2_2$. The
boundary term in (9) vanishes. Letting the two boundaries of
$M^\prime$ approach the two horizons, one finds the entropy or
action satisfies (7) as in the vacuum model.

For the electric case, the gauge field is
\begin{equation}
F =  \frac{iQ}{r^2}d \tau \wedge dr
\end{equation}
and the gauge potential is
\begin{equation}
A = i\left (\frac{Q}{r} - \Phi \right ) d\tau,
\end{equation}
where $\Phi$ is a constant to regularize the potential at the
horizon. One should regularize it separately for $M_l$ and $M_k$.

For this case, the boundary term in (9) takes the value $\beta
Q\Omega (\Phi_l -
\Phi_k)/4\pi$ where $\Phi_i = Q/r_i$ is the electric potential at
the horizon $r_i$, and $\Omega$ is the total solid angle of the
horizon. Here the multiple of the potential difference at the
two horizons $\Omega(\Phi_l - \Phi_k)/4\pi$ is interpreted as the
chemical potential associated with the electric charge. To
recover the microcanonical ensemble before substituting the
action into the
path integral for the partition function, one has to drop this
term. Again the formula (7) is valid. This prescription has been
previously used to recover the duality of the creation
probability between the magnetic and electric black hole
creations
in quantum cosmology [17][18]. If one replaces the horizon $r_k$
by infinity $r \longrightarrow \infty$, then the argument remains
similar to
that for the vacuum case.

The relation between the entropy and topology for a regular
gravitational instanton  has been studied. For many well
known regular instantons it has been
confirmed case by case that the entropy is one eighth of the
Euler characteristic for the
4-dimensional manifold  times the horizon
area [19]. This, however, is only correct due to the coincidence
that the Euler characteristic of the 2-sphere space $(\theta,
\phi)$ is 2. In a more general setting, such as a topological
black hole, the Euler characteristic of the section
$d\Omega^2_2$ is no longer 2. This is the reason that the formula
in [19] does not apply to a topological black hole [20]. In
contrast, formula (7) is true for all these cases mentioned.
Indeed, the entropy is connected
with the topology of the $(\tau, r)$ space only. The contribution
of the space $d\Omega^2_2$ to it is through the horizon areas.

Our analysis can be generalized into the Lovelock theory of
gravitation [21]. For an $n$-dimensional spacetime, the action
takes the form,
\begin{equation}
I =- \frac{1}{16\pi} \sum_{m =0}^{[(n-1)/2]}c_m(V_m + B_m),
\end{equation}
where the volume term $V_m$ is
\begin{equation}
V_m = \frac{1}{2^m} \int_M\delta^{a_1b_1\cdots
a_mb_m}_{c_1d_1 \cdots
c_md_m}R^{c_1d_1}_{a_1b_1}\cdots
R^{c_md_m}_{a_mb_m},
\end{equation}
where $\delta^{a_1b_1\cdots
a_mb_m}_{c_1d_1 \cdots c_md_m}$ is the generalized Kronecker
delta, and $[(n-1)/2]$ denotes the integer part of $(n-1)/2$. The
terms
$m=1$
and $m=0$ represent the Hilbert-Einstein action and the
cosmological constant term, respectively.

The boundary term $B_m$ takes the form [22]
\begin{equation}
B_m =-\frac{32 \pi}{n-2m} \oint_{\partial M} \pi^{ij}_m h_{ij},
\end{equation}
where $\pi^{ij}_m$ is the $m-$th component of the  momentum
conjugate to the metric $h_{ij}$
of  the boundary 
\begin{equation}
\pi^i_{j,m} = \sqrt{h} \sum_{s = 0}^{m-1}d_{s,m}
\delta^{a_1\cdots a_{2s}\cdots
a_{2m-1}i}_{b_1 \cdots b_{2s}\cdots
b_{2m-1}j}\bar{R}^{b_1b_2}_{a_1a_2}\cdots
\bar{R}^{b_{2s-1}b_{2s}}_{a_{2s-1}a_{2s}}K^{b_{2s+1}}_{a_{2s+1}}
\cdots
K^{b_{2m-1}}_{a_{2m-1}},
\end{equation}
where $\bar{R}^{ij}_{kl}$ are the curvature components related to
$\partial M$,  $K^i_j$ is the
extrinsic curvature , and 
\begin{equation}
d_{s,m} = \frac{(-1)^{m-s}2^{m-2s -5}m!}{\pi s! [2(m-s)-1]!!}.
\end{equation}

One assumes that the black hole takes the metric form (3) with
the metric $d\Omega^2_2$ replaced
by its $(n-2)$-dimensional counterpart $d\Omega^2_{n-2}$. 
Following the same argument, one finds that the action  of a
black hole  originates only from the neighbourhoods $M_i$ of
horizons of
the constrained instanton (to be justified below). The instanton
is obtained by identifying the imaginary time coordinate
using period $\beta$ as before. 

The curvature components are
\begin{equation}
R_{ijkl} = (\delta_{ik}\delta_{jl} - \delta_{il}\delta_{jk})
\frac{p - \Delta(r)}{r^2},
\end{equation}
where $i,j,k,l$ are indices in an orthonormal frame for the
metric $r^2d\Omega^2_{n-2}$, $p
= 1, 0, -1$ are for
the metric $d \Omega^2_{n-2}$ of the unit $(n-2)$-dimensional
sphere,
compactified   plane and compactified  hyperboloid, respectively.
As the boundary $\partial M_i$
approaches the
horizon,  all the extrinsic curvature components except
$K^{\tau}_{\tau}$ tend to zero. For curvature components (18) at
this limit, one has $\bar{R}_{ijkl}= R_{ijkl}$. 

The integral in  (15) for $\partial M_i$ can be factorized into
two parts. 
\begin{equation}
B_m = \frac{1}{2\pi}\oint Nd\tau K^{\tau}_{\tau}\cdot 4\pi m
\oint_{\partial
\bar{M}_i}V_{m-1}(\tilde{h}),
\end{equation}
where the first factor of the integral  is identified as the
second term of (6). If the parameter $\beta$ is not equal to
$2\pi \kappa_i$, where $\kappa_i$ is the
surface gravity, then a conical singularity emerges. In that
case, one has to add the third term
$\delta_i/2\pi$ of (6) here. If the horizon recedes into an
internal infinity, then this is no longer of concern. The volume
term corresponding to the first term in (6) tends to zero. By the
same argument, this factor
is the Euler characteristic $\chi(i)$
regardless of the existence of the singularity. The second
integral is over the $(n-2)$-dimensional
space $\partial \bar{M}_i$ of a section $\tau =
const.$ of the boundary  $\partial M_i$. All tilded
quantities are projected into the space  $\partial \bar{M}_i$ of
those in $\partial M_i$.

Using (18)  (19), one obtains the action $I_i$
for $M_i$
\begin{equation}
I_i = - \frac{\chi(i)A_i}{4} \sum_{m=1}^{[(n-1)/2]} \frac{mc_m
p^{m-1}(n-2)!}{(n-2m)!}
r_i^{-2m+2} \equiv - \frac{\chi(i)A_if_i}{4},
\end{equation}
where $A_i$ is the horizon area, and we define $0^0 = 1$ here.

The total action consists of those of the horizon neighbourhoods
for the closed or open black hole.
From our discussion, we have shown that the action is stationary,
and therefore the pasted
manifold is a constrained instanton. The entropy is the negative
of the action
\begin{equation}
S = - I =  \frac{1}{4} ( \chi(l) A_lf_l + \chi(k) A_kf_k).
\end{equation}

The discussion can be extended straightforward to the
nonvacuum  model. 
The  formula for the entropy of a black
hole with a regular instanton has been previously obtained
[23][24][14]. The
function $\Delta (r)$ for the Schwarzschild-like solution (3) may
not be a rational function [24][25][26][27][28], but our argument
for a black hole with distinct surface gravities is still
applicable.
 
The constrained instanton can be used as a seed for quantum
creation of a black hole in Lovelock
gravity. However, one has to use the instanton with the largest
action [8]. The relative
creation probability is the exponential of the entropy of the
instanton. After we have obtained 
formulas (7) and (21), we no longer need to check stationary
property of the constrained
instanton case by case for black hole creations in both the
Einstein and Lovelock gravities.

In summary, one can use microcanonical ensemble to calculate the
entropy of a black hole with distinct surface gravities. The
dominant contribution is due to the constrained instanton. The
instanton is justified by using the dimensional continuation of
the Gauss-Bonnet theorem. The quite universal formula (7) for the
entropy and its higher-dimensional version  (21) have been
proven. The entropy and quantum creation of a higher-dimensional
black hole in Lovelock theory are studied.

\vspace*{0.3in}
                       
\bf References:

\vspace*{0.1in}
\rm

1. G.W. Gibbons and S.W. Hawking, \it Phys. Rev. \bf D\rm
\underline{15}, 2725 (1977).

2. G.W. Gibbons and S.W. Hawking, \it Phys. Rev. \bf D\rm
\underline{15}, 2738 (1977).

3. S.W. Hawking, G.T. Horowitz and S.F. Ross, \it Phys. Rev. \bf
D\rm\underline{51}, 4302 (1995).

4. J.B. Hartle and S.W. Hawking, \it Phys. Rev. \rm \bf D\rm
\underline{28}, 2960 (1983).

5. R. Bousso and S.W. Hawking, \it Phys. Rev. \rm \bf D\rm
\underline{52}, 5659 (1995), gr-qc/9506047.

6. Z.C. Wu, \it Gene. Relativ. Grav. \rm\underline{30}, 1639
(1998), hep-th/9803121.

7. Z.C. Wu, \it Int. J. Mod. Phys. \rm \bf D\rm\underline{6}, 199
(1997), gr-qc/9801020.

8. Z.C. Wu, \it Phys. Lett. \rm \bf B\rm
\underline{445}, 274 (1998), gr-qc/9810012.

9. Z.C. Wu, gr-qc/9907064.

10. Z.C. Wu, gr-qc/9907065.

11. C. Teitelboim, \it Phys. Rev. \rm \bf D\rm
\underline{51}, 4315 (1995).

12. S.W. Hawking and G.T. Horowitz, \it Class. Quant. Grav. \rm
\rm\underline{13}, 1487 (1996). 

13. T. Regge and C. Teitelboim, \it Ann. Phys. \rm (N.Y.)
\underline{88}, 286 (1974).

14. M. Ba$\tilde{n}$\rm ados, C. Teitelboim and J. Zanelli, 
\it Phys. Rev. Lett. \bf \rm \underline{72}, 957 (1994).

15. G.W. Gibbons and R.E. Kallosh, \it Phys. Rev. \rm \bf D\rm
\underline{51}, 2839 (1995).

16. J.D. Brown, E.A. Martinez and J.W. York, \it Phys. Rev. Lett.
\bf \rm \underline{66}, 2281 (1991).

17. S.W. Hawking and S.F. Ross,  \it Phys. Rev. \bf D\rm 
\underline{52}, 5865 (1995).

18. R.B. Mann and S.F. Ross, \it Phys. Rev. \bf D\rm
\underline{52}, 2254 (1995).  

19. S. Liberati and G. Pollifrone, \it Phys. Rev. \bf D\rm
\underline{56}, 6458 (1997). 

20. D.R. Brill, J. Louko and P. Peldan, \it Phys. Rev. \rm \bf
D\rm \underline{56}, 3600 (1997).

21. D. Lovelock, \it J. Math. Phys. \rm \underline{12}, 498
(1971).

22. C. Teitelboim and J. Zanelli, \it Class. Quant. Grav. \rm
\underline{4}, L125 (1987).

23. T. Jacobson and R. Myers,  \it Phys. Rev. Lett. \bf \rm
\underline{70}, 3684 (1993).

24. R.C. Myers and J.Z. Simon, \it Phys. Rev. \rm \bf D\rm
\underline{38}, 2434 (1988).

25. D.G. Boulware and S. Deser,  \it Phys. Rev. Lett. \bf \rm
\underline{55}, 2656 (1985).

26. J.T. Wheeler,  \it  Nucl. Phys. \bf B\rm
\underline{273}, 732 (1986).

27. R.C. Myers and M.J. Perry, \it Ann. Phys. \rm (N.Y.)
\underline{172}, 304 (1986).

28. D.L. Wiltshire, \it Phys. Rev. \rm \bf D\rm
\underline{38}, 2445 (1988).

\end{document}